\documentclass[cameraready]{Interspeech}

\newcommand{\benchname}{\textsc{ParaPairAudioBench}}
\title{\benchname: Paralinguistic Pairwise Audio Benchmark for \audiollmjudge}

\author[affiliation={1 \clubsuit}]{Jisu}{Jeon}
\author[affiliation={2 \spadesuit}]{Seungyeon}{Jwa}
\author[affiliation={3}]{Joosung}{Lee}
\author[affiliation={3}]{Jinhyeon}{Kim}
\author[affiliation={4 \clubsuit}]{Woojin}{Chung}
\author[affiliation={3}]{Hwiyeol}{Jo}
\author[affiliation={3,4}]{Jeonghoon}{Kim}
\author[affiliation={2}]{Jonghyun}{Choi}
\author[affiliation={3 \dagger}]{Soyoon}{Kim}

\address{
    $^1$ Hongik University, South Korea \\
    $^2$ Seoul National University, South Korea \\
    $^3$ NAVER Cloud, South Korea \\
    $^4$ KAIST, South Korea 
}

\email{first@university.edu, second@companyA.com, third@companyB.ai}
\email{wltnjeon0119@gmail.com,soyoon.kim@navercorp.com}

\usepackage{xcolor}
\usepackage{comment}
\usepackage{kotex}
\usepackage{booktabs}
\usepackage{tabularx}
\usepackage{multirow}
\usepackage{makecell}
\usepackage{threeparttable}
\usepackage{array}
\newcolumntype{Y}{>{\centering\arraybackslash}X}
\usepackage{todonotes}
\usepackage{colortbl}
\usepackage{makecell}
\usepackage{float}
\usepackage{graphicx}
\usepackage{pgfplots}
\pgfplotsset{compat=1.18}
\newcommand{\audiollm}{LALM}
\newcommand{\audiollms}{LALMs}
\newcommand{\tie}{\texttt{\small [[Tie]]}}

\newcommand{\audiollmjudge}{LALM-as-a-Judge}

\newcommand{\blfootnote}[1]{%
  \begingroup
  \renewcommand\thefootnote{}%
  \footnote{#1}%
  \addtocounter{footnote}{-1}%
  \endgroup
  \ignorespaces
}

\begin{document}

\maketitle
\begin{abstract}
Large Audio-Language Models (LALMs) have been widely used as judge models for the automatic evaluation of generated speech.
However, prior approaches predominantly focus on holistic naturalness, leaving fine-grained paralinguistic distinctions underexplored.
We introduce \benchname{}, a pairwise benchmark of 5,175 audio pairs across five paralinguistic dimensions: Style, Rate, Emphasis, Age, and Gender.
Our experiments show that current LALM judges still lag behind human judgments by 32\%p on average and exhibit severe calibration failures, particularly in Tie cases where the correct decision is to abstain.
To further analyze lexical versus acoustic reliance, the benchmark includes both same-transcript and cross-transcript conditions.
\benchname{} enables multi-dimensional, calibration-aware assessment of the reliability of LALM-as-a-Judge for paralinguistic speech evaluation.
\end{abstract}

\section{Introduction}

Speech generation systems have achieved remarkable progress, producing highly natural and expressive speech across diverse speakers and styles \cite{seedtts, naturalspeech3, cosyvoice2, t5tts}. 
However, naturalness alone does not guarantee that generated speech correctly follows the intended paralinguistic cues such as speaking rate, emphasis, or speaker characteristics. 
As a result, evaluation methodology has not kept pace with these advances, and assessing whether synthesized speech truly meets the fine-grained perceptual requirements remains an open challenge \cite{instructttseval,emergenttts,paras2s}.
\blfootnote{\textsuperscript{$\clubsuit$}Work done during an internship at NAVER Cloud.}
\blfootnote{\textsuperscript{$\spadesuit$}Work done during a residency program at NAVER Cloud.}
\blfootnote{\textsuperscript{$\dagger$}Corresponding author.}
\begin{figure}[t]
  \centering  \includegraphics[width=1\columnwidth]{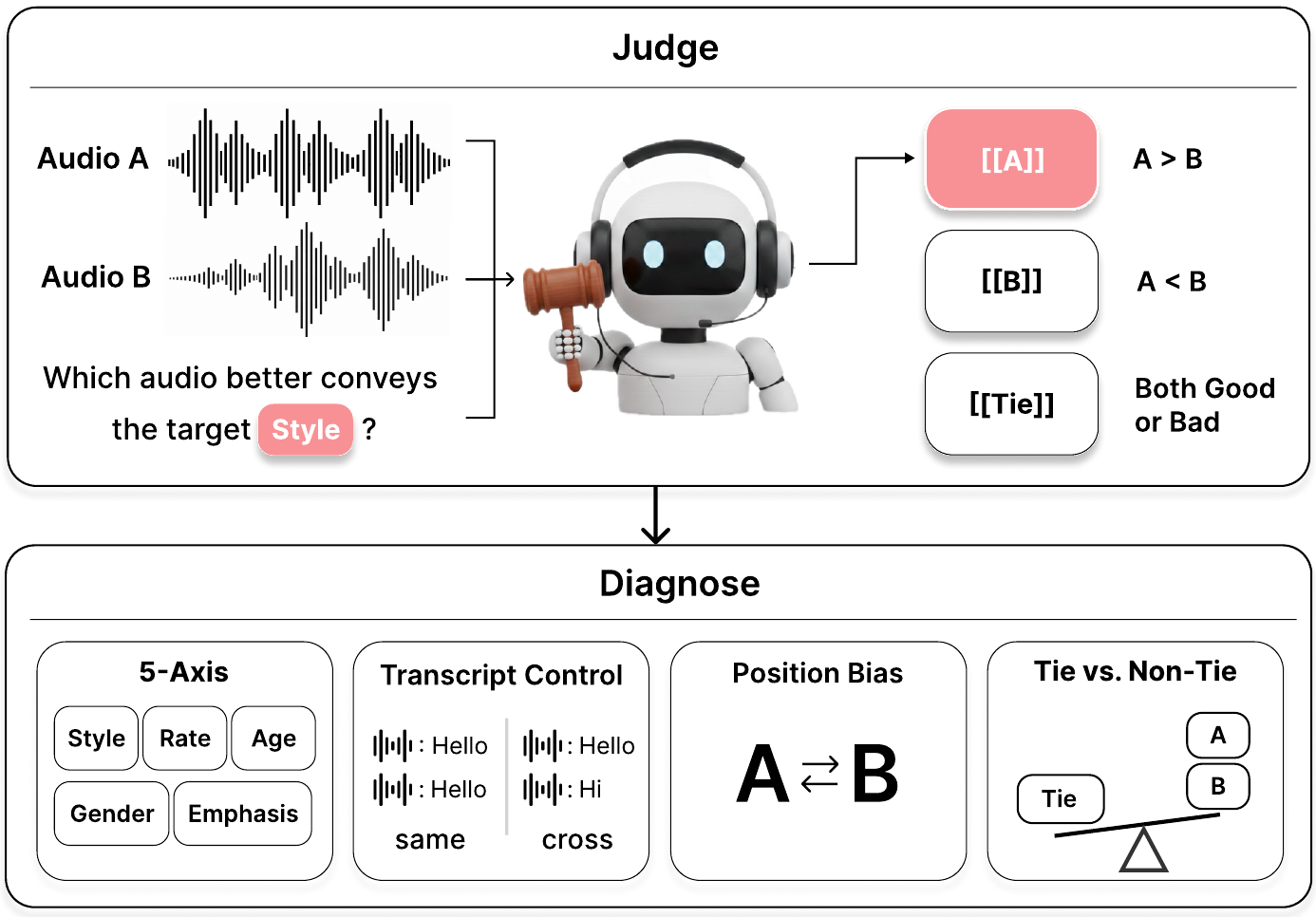}
\caption{
\textbf{Overview of \benchname{}.} Given a pair of speech samples and an instruction specifying the evaluation criterion, \audiollmjudge{} assesses them and outputs one of \textsc{[[A]]}, \textsc{[[B]]}, or \textsc{[[Tie]]}.
}
\label{fig:overview}
\end{figure}

To address this challenge, we introduce \benchname{}\footnote{\href{https://github.com/jsujeon/ParaPairAudioBench}{Official Data and Code : https://github.com/jsujeon/ParaPairAudioBench}}, a diagnostic pairwise benchmark comprising 5,175 audio pairs across five paralinguistic dimensions: Style, Rate, Emphasis, Age, and Gender. We adopt a pairwise comparison format, as pairwise judgments reduce the calibration burden on the judges and have been shown to yield more reliable preferences in both text \cite{zheng2023judging,alpacaevals} and speech \cite{speechjudge} evaluation. 
Each instance allows three possible decisions: Audio A, Audio B, or \textit{Tie}, which represents a draw when both samples are equally appropriate for the target criterion. The benchmark therefore includes both the \textit{Tie} cases, where a draw is possible, and \textit{Non-Tie} cases, where one sample must be preferred over the other.

We evaluate several representative LALMs, including both commercial and open-source systems. The results show that current LALMs still lag behind human performance and exhibit systematic weaknesses across different paralinguistic criteria. 
Our analysis further reveals systematic behaviors: strong performance on globally distributed cues like speech rate, but clear weaknesses on localized prosodic attributes such as emphasis. 
These findings suggest that fine-grained and diagnostic evaluation is necessary to better understand the strengths and limitations of LALM-as-a-Judge systems.

Our primary contributions are as follows:
\begin{itemize}
    \item \textbf{Diagnostic Multi-dimensional Evaluation:} We decompose paralinguistic judgment into five distinct criteria, revealing that models excel at globally distributed cues such as Rate but struggle with localized prosodic prominence in Emphasis, a finding invisible to the aggregate naturalness.
    \item \textbf{Calibration and Robustness Assessment:} We use \textit{calibration} to refer to the judge's ability to abstain via \tie{} under ambiguity. We incorporate explicit \tie{} conditions and presentation-order swaps, exposing systematic calibration failure and position bias of up to 29.4\%p.
    \item \textbf{Textual vs.\ Acoustic Dependency Analysis:} By controlling  transcript matching, we find that models over-rely on textual cues for Style while overlooking prosodic context for Emphasis, demonstrating the need for modality-aware evaluation.
\end{itemize}

\section{Related Works}

To automate evaluation while preserving human-like judgment, the \textit{LLM-as-a-Judge} paradigm \cite{zheng2023judging, alpacaevals, prometheus2, chatbotarena} has shown strong alignment with human preferences in the text domain. However, since this paradigm operates primarily on textual representations, it is unable to directly evaluate the acoustic and paralinguistic properties of speech.
To address this, recent work has explored \audiollms{} as automated speech quality judges. SpeechJudge \cite{speechjudge} fine-tunes an \audiollm{} to produce naturalness-based pairwise preferences, while AudioJudge \cite{audiojudge} prompts an \audiollm{} to evaluate holistic speech quality. 
Subsequent studies have further analyzed the behavior of models as judges, investigating their decision processes through interpretable explanations \cite{chiang2025audioawarelargelanguagemodels,speechllmasjudge, chen2025audiolargelanguagemodels}.
Beyond overall speech quality, emerging research has expanded to evaluate paralinguistic and expressive aspects of speech generation, including speaking style, prosody, and other vocal attributes \cite{emphassess, liu2025vocalbench}. 
These studies apply model-as-a-judge pipelines to assess expressive TTS scenarios and decompose speech-to-speech evaluation into multiple perceptual dimensions such as content fidelity and paralinguistic attributes \cite{emergenttts,trace}. 
Other benchmarks further assess audio–language understanding or speech-to-speech interaction through automated evaluation \cite{paras2s,sakura}.


Despite these advances, the reliability of \audiollmjudge{} under controlled pairwise evaluation settings remains largely unexplored. 
Prior work has primarily focused on overall preference judgments or audio-language understanding tasks, with limited analysis of how reliably \audiollms{} evaluate paralinguistic attributes under controlled conditions. 
To facilitate analysis of \audiollmjudge{} reliability, we introduce \benchname{}, a diagnostic pairwise benchmark that factorizes paralinguistic judgment into criterion-level discrimination, Tie calibration, transcript sensitivity, and order robustness.

\section{Benchmark Design}

\begin{table}[t]
\centering
\footnotesize
\setlength{\tabcolsep}{4pt} 
\renewcommand{\arraystretch}{1.15}
\begin{tabular}{@{}lccccc|c@{}}
\toprule
\textbf{Criterion} & \textbf{Style} & \textbf{\begin{tabular}[c]{@{}c@{}} Rate\end{tabular}} & \textbf{Emphasis} & \textbf{Age} & \textbf{Gender} & \textbf{Total} \\
\midrule
Tie         & 720   & --    & 445   & 750   & 500   & 2,415 \\
Non-Tie     & 720   & 375   & 415   & 750   & 500   & 2,760 \\
\midrule
\textbf{Total} & \textbf{1,440} & \textbf{375} & \textbf{860} & \textbf{1,500} & \textbf{1,000} & \textbf{5,175} \\
\bottomrule
\end{tabular}
\caption{Dataset statistics of \benchname{}.}
\label{tab:dataset_stats}
\end{table}

\subsection{Overview}

\benchname{} is a controlled diagnostic pairwise benchmark across five paralinguistic criteria---Style, Rate, Emphasis, Age, and Gender---designed to diagnose whether a judge model can consistently identify which sample better satisfies a given criterion. The dataset comprises 5,175 instances from the validated public corpora: Expresso for Emphasis and Style \cite{expresso}, Sonos Voice Control Bias Assessment for Age and Gender \cite{svc}, LibriTTS for Gender \cite{libritts}, and EARS for Rate \cite{ears}. To mitigate potential pretraining contamination, we use only the official test splits and retain pairs that satisfy criterion-specific label constraints, transcript-control requirements, and balancing conditions where applicable. Dataset statistics are shown in Table~\ref{tab:dataset_stats}.
    
\subsection{Selection of Criteria}

The five criteria were selected to span the hierarchy of speech variation, from static speaker traits to dynamic prosodic controls. Gender and Age represent stable physiological characteristics grounded in vocal tract resonance and fundamental frequency. Rate and Emphasis target the dynamic prosodic variations: Rate reflects the global temporal pacing across the utterance, while Emphasis captures the localized prominence shifts at word level through the changes in energy, pitch, and duration. Style serves as a high-level expressive attribute spanning from acoustic texture, such as whispering, to affective intents, such as happiness or sadness. Together, these five criteria cover both the static speaker characteristics (Gender, Age) and the dynamic prosodic controls (Style, Rate, Emphasis), enabling us to identify where LALM judges succeed and where they fail. Concretely, each criterion defines a pairwise judgment task in which the judge determines which utterance better fulfills a given speech generation instruction: for \textbf{Style}, speaking in a specified manner (e.g., whispering, confused); for \textbf{Rate}, speaking at a desired pace; for \textbf{Emphasis}, placing stress on a designated word; for \textbf{Age}, sounding like a specified age group (e.g., child, elderly); and for \textbf{Gender}, sounding like a specified gender (e.g., male, female).

\subsection{Design Principles}

The benchmark design is guided by two objectives: exposing forced preference under ambiguity through tie conditions, and isolating lexical cues from acoustic cues. Pairs are subsampled from the official test splits and stratified by criterion-specific labels where available. Non-Tie pairs contrast a target sample with a non-target sample, while Tie pairs contain two samples with the same target compatibility, as detailed below. Since pair construction must jointly satisfy these label, transcript, and balancing constraints, we prioritize controlled diagnostic cells over exhaustive enumeration of corpus pairs.

\textbf{Tie conditions.}\quad To evaluate whether models can abstain under ambiguity rather than forcing a preference, we incorporate explicit tie conditions: models may choose \tie{} when no meaningful distinction exists. For Style, Age, and Gender, tie cases are balanced between ``Both Good'' cases where both samples satisfy the target and ``Both Bad'' cases where neither satisfies it; in either scenario, \tie{} is the correct response. For Emphasis, ties consist only of ``Both Bad'' instances: because emphasis is a localized prominence phenomenon, constructing ``Both Good'' ties would require two samples with perceptually identical stress profiles on the same word, which cannot be reliably verified without frame-level annotation. Rate excludes ties entirely, as naturally recorded speech exhibits rhythmic variation even under identical rate labels; we leave rate-level tie construction for future work.

\textbf{Transcript control.}\quad The key advantage of LALM judges over text-based LLMs lies in their ability to process acoustic signals. To examine whether models genuinely leverage this acoustic capability or simply rely on lexical content, the pairs are balanced between same-transcript (47.0\%) and cross-transcript (53.0\%) conditions. This transcript-controlled comparison preserves the model's audio-processing pathway while holding or varying lexical content, allowing us to diagnose acoustic versus ASR-recoverable textual reliance. Rate pairs use only identical transcripts to avoid sentence-length confounds. For Emphasis, cross-transcript pairs are constructed such that both sentences contain the target word, ensuring that Emphasis judgment is not confounded by the absence of the lexical target.


\definecolor{lightpink}{RGB}{255,204,204}
\definecolor{lightgreen}{RGB}{204,255,204}

\begin{table}[t]
\centering
\scriptsize
\setlength{\tabcolsep}{2pt}
\renewcommand{\arraystretch}{1.05}
\begin{tabular}{l *{5}{>{\centering\arraybackslash}m{0.75cm}} c}\toprule
\textbf{Model / Criterion} & \bf Style & \bf Rate & \bf Emph. & \bf Age & \bf Gender & \bf Avg. \\
\midrule
Human\textsuperscript{\dag} 
& \cellcolor{lightgreen}85.7 
& \cellcolor{lightgreen}91.0 
& \cellcolor{lightgreen}85.7 
& \cellcolor{lightgreen}52.7 
& \cellcolor{lightgreen}80.7 
& 79.2 \\
\midrule
Gemini 2.5 Flash 
& \cellcolor{lightgreen}\textbf{48.5}
& \cellcolor{lightgreen}\textbf{88.9} 
& \cellcolor{lightgreen}\textbf{49.7}
& \cellcolor{lightgreen}\textbf{\underline{56.5}} 
& \cellcolor{lightgreen}\textbf{64.2} 
& \textbf{61.5} \\
GPT-4o Audio 
& \cellcolor{lightgreen}36.4 
& \cellcolor{lightgreen}77.6 
& \cellcolor{lightgreen}43.8 
& \cellcolor{white}34.9 
& \cellcolor{lightgreen}39.3 
& 46.4 \\
SpeechJudge-7B 
& \cellcolor{white}32.6 
& \cellcolor{white}48.0 
& \cellcolor{white}32.9 
& \cellcolor{lightpink}25.8 
& \cellcolor{lightpink}29.9 
& 33.8 \\
Kimi-Audio-7B 
& \cellcolor{lightgreen}45.9
& \cellcolor{lightgreen}76.0 
& \cellcolor{lightgreen}42.9 
& \cellcolor{lightpink}27.5 
& \cellcolor{lightgreen}58.6 
& 50.2 \\
Qwen2.5-Omni-7B 
& \cellcolor{white}35.8 
& \cellcolor{lightgreen}61.9 
& \cellcolor{lightgreen}36.7 
& \cellcolor{lightgreen}38.1 
& \cellcolor{lightgreen}47.4 
& 44.0 \\
\bottomrule
\end{tabular}
\caption{
\textbf{Overall accuracy (\%) on \benchname{}.}
\textsuperscript{\dag}Human: n=50 per criterion. \colorbox{lightgreen}{Green}/\colorbox{lightpink}{pink}: above/below chance (z-test, 95\% confidence level). \textbf{Bold}: best model per criterion.
\underline{Underlined}: higher than human.
}
\label{tab:overall_acc_onecol}
\end{table}

\section{Experimental Setup}

\subsection{Input Format}

Each model receives two audio inputs, Audio A and Audio B, along with an instruction specifying the target criterion regarding the given attribute, and is required to select the audio that better satisfies the criterion. If no meaningful distinction can be made, the model must select \tie{}. When presenting two audio inputs, we provide Audio A and Audio B as separate audio segments within the same inference call, explicitly marked by textual identifiers (“Audio A” and “Audio B”). \footnote{We also experimented with concatenating the two waveforms into a single continuous audio stream following AudioJudge \cite{audiojudge}, but observed no substantial difference in accuracy.} To ensure fair comparison, we restrict our evaluation to \audiollms{} capable of processing multiple audio inputs within a single inference call. Throughout this paper, \textit{chance level} refers to the expected accuracy of a uniform random baseline: 33.3\% for three-way criteria and 50\% for Rate.

\subsection{Evaluation Protocol and Baselines}

We evaluate five models representing three categories: two commercial models, Gemini 2.5 Flash and GPT-4o Audio; two open-source models, Kimi-Audio-7B-Instruct and Qwen2.5-Omni-7B; and one fine-tuned judge model, SpeechJudge-7B \cite{gemini25, gpt4o, kimiaudio, qwen25omni, speechjudge}.
For each evaluation criterion, we use a criterion-specific prompt template. The same template is applied to all models without modification to ensure fair comparison. Since some \audiollms{} incorporate ASR-pretrained encoders, their judgments may rely not only on the acoustic cues but also on the implicitly transcribed textual information. To analyze this effect, we separately measure the accuracy under the same- and cross-transcript conditions. 
To account for the position bias, each pair is evaluated twice by swapping the input order. Specifically, Acc@A denotes the accuracy when all non-tie pairs are arranged so that the correct answer is always in position A, and Acc@B denotes the accuracy when the correct answer is always in position B; in both cases, accuracy is computed over all instances including ties (Rate is excluded, as it contains no tie cases). The gap between Acc@A and Acc@B thus directly quantifies the position bias.
Overall accuracy is the average of the two swap conditions, and we report Non-Tie and Tie accuracy separately. SpeechJudge follows its original protocol of majority voting over 10 inference runs with temperature 1.0 \cite{speechjudge}; all other models use greedy decoding. 

\textbf{Human evaluation.}\quad 
We report human performance on a 250-item subset (50 per criterion), balanced as closely as possible across Tie/Non-Tie, transcript, and answer-position conditions. Six raters independently evaluated each pair; A/B-swapped duplicates were omitted to avoid memory contamination, leaving order invariance to model-side evaluation. Inter-rater reliability was Fleiss’ $\kappa = 0.67$ \cite{fleiss1971measuring,landis1977measurement}.

\begin{table}[t]
\centering
\scriptsize
\setlength{\tabcolsep}{3.1pt}
\renewcommand{\arraystretch}{1.1}
\begin{tabular}{l | cc | cc | cc | cc}
\toprule
& \multicolumn{2}{c|}{\textbf{Style}} & \multicolumn{2}{c|}{\textbf{Emph.}} & \multicolumn{2}{c|}{\textbf{Age}} & \multicolumn{2}{c}{\textbf{Gender}} \\
\textbf{Model / Criterion} & \textbf{NT} & \textbf{T} & \textbf{NT} & \textbf{T} & \textbf{NT} & \textbf{T} & \textbf{NT} & \textbf{T} \\
\midrule
Human\textsuperscript{\dag}  & \cellcolor{lightgreen}86.0 & \cellcolor{lightgreen}85.3 & \cellcolor{lightgreen}95.3 & \cellcolor{lightgreen}76.0 & \cellcolor{lightgreen}66.7 & \cellcolor{lightgreen}38.7 & \cellcolor{lightgreen}86.0 & \cellcolor{lightgreen}75.3 \\
\midrule
Gemini 2.5 Flash & \cellcolor{lightgreen}79.0 & \cellcolor{lightpink}\textbf{18.0} & \cellcolor{lightgreen}\textbf{82.3} & \cellcolor{lightpink}\textbf{19.3} & \cellcolor{lightgreen}\textbf{\underline{77.5}} & \cellcolor{white}35.5 & \cellcolor{lightgreen}\textbf{84.9} & \cellcolor{lightgreen}43.4 \\
GPT-4o Audio & \cellcolor{lightgreen}69.0 & \cellcolor{lightpink}3.8 & \cellcolor{lightgreen}74.5 & \cellcolor{lightpink}15.3 & \cellcolor{lightgreen}61.0 & \cellcolor{lightpink}8.7 & \cellcolor{lightgreen}65.3 & \cellcolor{lightpink}13.2 \\
SpeechJudge-7B & \cellcolor{lightgreen}63.4 & \cellcolor{lightpink}1.7 & \cellcolor{lightgreen}63.0 & \cellcolor{lightpink}4.7 & \cellcolor{lightgreen}47.7 & \cellcolor{lightpink}3.9 & \cellcolor{lightgreen}58.2 & \cellcolor{lightpink}1.6 \\
Kimi-Audio-7B & \cellcolor{lightgreen}\textbf{81.0} & \cellcolor{lightpink}10.7 & \cellcolor{lightgreen}79.3 & \cellcolor{lightpink}9.0 & \cellcolor{lightpink}26.3 & \cellcolor{lightpink}28.7 & \cellcolor{lightgreen}66.7 & \cellcolor{lightgreen}50.4 \\
Qwen2.5-Omni-7B & \cellcolor{lightgreen}54.0 & \cellcolor{lightpink}17.6 & \cellcolor{lightgreen}65.1 & \cellcolor{lightpink}10.2 & \cellcolor{lightgreen}38.0 & \cellcolor{lightgreen}\textbf{38.2} & \cellcolor{lightgreen}39.9 & \cellcolor{lightgreen}\textbf{54.9} \\
\bottomrule
\end{tabular}
\caption{\textbf{Non-Tie (NT) vs.\ Tie (T) accuracy (\%).} Rate omitted (no Tie cases). \textsuperscript{\dag}Human: n=50 per criterion. \colorbox{lightgreen}{Green}/\colorbox{lightpink}{pink}: above/below chance (z-test, 95\% confidence level). \textbf{Bold}: best model per criterion. \underline{Underlined}: higher than human.}
\label{tab:nt_vs_t_final}
\end{table}
\section{Analysis}

\subsection{Overall Performance and Human--Model Gap}


Table~\ref{tab:overall_acc_onecol} reveals three key patterns. First, even the strongest model, Gemini 2.5 Flash, falls 17.7\%p below human performance on average. Second, excluding Rate---which involves only binary choices---models on average achieve their highest performance on Gender and Emphasis, whereas humans perform best on Emphasis and Style. This contrast suggests that models and humans rely on different cues when evaluating paralinguistic attributes. Third, Age shows relatively low accuracy even for human raters, reflecting the inherent difficulty of estimating age from speech. Gemini 2.5 Flash slightly surpasses human performance on this criterion by 3.8\%p. The inter-rater agreement for Age is also the lowest among all criteria ($\kappa=0.365$), indicating that age perception from speech is challenging even for human listeners.

\subsection{Calibration Failure: Non-Tie vs.\ Tie}

Table~\ref{tab:nt_vs_t_final} reveals a pervasive calibration failure: models systematically force a preference rather than recognizing indeterminate cases. Gemini 2.5 Flash achieves 82.3\% on Non-Tie Emphasis cases but drops to 19.3\% on Tie, and GPT-4o Audio records 69.0\% on Non-Tie Style cases but only 3.8\% on Tie. SpeechJudge shows the most extreme collapse, with Tie accuracy at 1.7\% on Style and 1.6\% on Gender, almost never selecting Tie.

This calibration gap varies by criterion. Emphasis and Style exhibit the most severe collapse, with Tie accuracy below 20\%, indicating that models cannot abstain when the discriminative cues are expressive or prosodic. Gender shows a mixed pattern, where some models retain moderate Tie accuracy while others almost never select Tie. Humans achieve over 75\% Tie accuracy on Emphasis, Gender, and Style, confirming that the acoustic cues in these conditions are perceptually sufficient for abstention.
A different pattern appears for Age. Gemini 2.5 Flash achieves strong Non-Tie accuracy that even exceeds human performance, yet its Tie accuracy remains slightly lower than the human level. This suggests that models can detect large age differences but struggle to abstain when age cues are ambiguous, highlighting the calibration difficulty in graded perceptual attributes.
Open-source models like Kimi-Audio and Qwen2.5-Omni show an inverted pattern on Age, where Tie accuracy exceeds Non-Tie. However, since their Non-Tie performance is already near or below \textit{chance level}, this inversion likely indicates that these models fail to perform the age discrimination task altogether rather than exhibiting meaningful calibration.

\subsection{Criterion-specific Findings}
\label{sec:criterion_findings}
Decomposing evaluation into five axes exposes behavioral patterns that a single naturalness score would obscure.
Gender, a categorical attribute with salient markers such as F0 and formant structure, shows the highest Non-Tie accuracy at 84.9\% for Gemini 2.5 Flash. However, Tie accuracy still drops to 43.4\%, demonstrating that strong discrimination does not guarantee calibration. Style presents a contrasting case: Gemini achieves a comparable Non-Tie score of 79.0\%, likely because global expressive cues such as laughter or whispering are perceptually salient. Yet Style exhibits the most severe Tie collapse and the largest transcript dependency, revealing that the high Non-Tie accuracy can mask deeper vulnerabilities, which will be detailed in Section~\ref{sec:transcript}.

\subsection{Transcript Influence: Acoustic vs.\ Lexical Reliance}
\label{sec:transcript}
\begin{table}[t]
\centering
\scriptsize
\setlength{\tabcolsep}{2.5pt}
\renewcommand{\arraystretch}{1.1}
\begin{tabular}{l l c | ccccc}
\toprule
& & \textbf{Human} & \textbf{Gemini} & \textbf{GPT-4o} & \textbf{Speech} & \textbf{Kimi-} & \textbf{Qwen2.5-} \\
& & & \textbf{2.5 Flash} & \textbf{Audio} & \textbf{Judge-7B} & \textbf{Audio-7B} & \textbf{Omni-7B} \\
\midrule

\multirow{2}{*}{Style}
& Same  & \cellcolor{lightgreen}86.0 & \cellcolor{lightgreen}83.8 & \cellcolor{lightgreen}66.2 & \cellcolor{lightgreen}65.6 & \cellcolor{lightgreen}83.6 & \cellcolor{lightgreen}44.6 \\
& Cross & \cellcolor{lightgreen}85.3 & \cellcolor{lightgreen}36.6 & \cellcolor{lightpink}26.4 & \cellcolor{lightpink}21.5 & \cellcolor{white}33.2 & \cellcolor{white}32.9 \\
\midrule
\multirow{2}{*}{Emph.}
& Same  & \cellcolor{lightgreen}87.3 & \cellcolor{lightgreen}43.5 & \cellcolor{white}34.9 & \cellcolor{white}30.8 & \cellcolor{lightgreen}40.8 & \cellcolor{lightgreen}39.2 \\
& Cross & \cellcolor{lightgreen}84.0 & \cellcolor{lightgreen}56.4 & \cellcolor{lightgreen}53.4 & \cellcolor{white}35.1 & \cellcolor{lightgreen}45.2 & \cellcolor{white}34.0 \\
\bottomrule
\end{tabular}
\caption{\textbf{Accuracy (\%) by transcript condition (chance: 33.3\%).}
\textit{Same}: identical transcripts; \textit{Cross}: different transcripts.\colorbox{lightgreen}{Green}/\colorbox{lightpink}{pink}: above/below chance (z-test, 95\% confidence level). }
\label{tab:transcript_gap}
\end{table}
Table~\ref{tab:transcript_gap} reveals a striking Style--Emphasis asymmetry, visible only through multi-axis, transcript-controlled evaluation.

For Style, all models perform dramatically better under the Same transcript condition. Gemini 2.5 Flash records 83.8\% on Same vs.\ 36.6\% on Cross, with a 47.2\%p gap, while humans maintain stable accuracy across both at 86.0\% and 85.3\%. When transcripts are identical, the lexical content is controlled and models can focus on the acoustic cues such as prosody, rhythm, and timbre. The sharp drop under the Cross transcript condition indicates that models rely heavily on the textual content for style judgment, and when lexical variation is introduced, it confounds the discrimination rather than aiding it.

For Emphasis, the pattern reverses. Gemini 2.5 Flash achieves 43.5\% on Same transcript but 56.4\% on Cross, and GPT-4o Audio shows a similar trend at 34.9\% vs.\ 53.4\%. Same-transcript pairs require detecting the subtle word-level acoustic variations in energy, pitch, and duration within the identical utterances. Cross-transcript pairs provide additional global prosodic scaffolding, where differences in sentence rhythm and intonation amplify perceived emphasis contrasts, compensating for the weak local sensitivity. Even in the Cross condition, both utterances still contain the same target word whose emphasis must be judged; only the surrounding sentence differs. Humans again remain stable across both conditions, confirming this is a model-specific limitation.

This Style--Emphasis contrast provides a central diagnostic finding: current \audiollms{} leverage global acoustic and lexical cues but struggle with localized prosodic discrimination. This distinction would be invisible under a single-axis evaluation, as both criteria would be collapsed into one aggregate score.

\subsection{Position Bias and Consistency}
\label{sec:position_bias}

\begin{figure}[t]
  \centering \includegraphics[width=1.05\columnwidth]{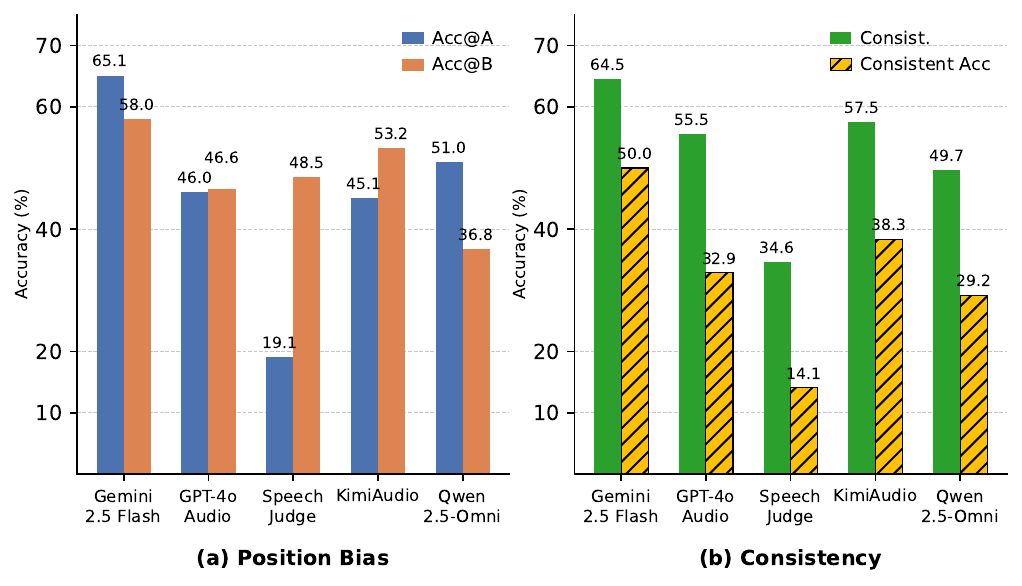}
\caption{
\textbf{Position bias and consistency under presentation order swaps.}
(a) Acc@A/Acc@B: accuracy when the correct audio appears first or second; the gap indicates positional bias.
(b) Consistency/ConsistentAcc: identical decisions under both orders; consistent and correct decisions.
}
\label{fig:position_bias}
\end{figure}

As shown in Figure~\ref{fig:position_bias}, all evaluated models exhibit systematic position bias in their judgments. SpeechJudge shows the most pronounced bias with an average Acc@A--Acc@B gap of 29.4\%p, consistently favoring position B. The direction of the bias also differs across models.
Qwen2.5-Omni consistently favors position A across all five criteria, whereas SpeechJudge consistently favors position B.

We define \textit{consistency} as the fraction of all pairs where the model makes the same decision (\textsc{[[A]]}/\textsc{[[B]]}/\textsc{[[Tie]]}) under both presentation orders, and \textit{ConsistentAcc} as the fraction of all pairs where that decision is both consistent and correct.
The model ranking is identical across overall accuracy, consistency, and consistent accuracy. Gemini 2.5 Flash consistently achieves the highest performance across these measures, whereas SpeechJudge remains the lowest. This pattern is consistent with the design of SpeechJudge, which is fine-tuned primarily for holistic naturalness assessment rather than attribute-specific paralinguistic comparison. As a result, its judgments may not transfer well to controllable paralinguistic attributes such as style, rate, emphasis, age, or gender.

\section{Conclusion}

We introduced \benchname{}, a pairwise benchmark of 5,175 audio pairs designed to diagnose the reliability of \audiollmjudge{} systems across five controllable paralinguistic criteria: Style, Rate, Emphasis, Age, and Gender. Our evaluation shows that current LALM judges still fall short of human judgments on average and exhibit systematic failure modes that are not visible under aggregate naturalness evaluation.
We observe a pervasive calibration failure in Tie cases, where models frequently force a preference even when abstention is the correct decision. Transcript-controlled analysis reveals asymmetric modality dependence: models over-rely on lexical cues for Style judgments, while Emphasis judgments benefit from cross-transcript prosodic context, indicating limited local prosodic sensitivity within identical utterances. Finally, presentation order swaps expose consistent positional bias and limited swap-robust performance, underscoring the need to evaluate order invariance in judge pipelines.
As a controlled benchmark built from public natural-speech corpora, \benchname{} thus prioritizes reliable diagnostic contrasts over exhaustive coverage of every possible condition. Future work can extend the protocol toward broader and more uniformly balanced coverage, together with debiasing methods for improving calibration and order robustness.


\section{Use of generative AI tools}
Generative AI tools were used for editing and polishing the text and for generating figures in this paper. All research design, experiments, analysis, and intellectual contributions are entirely the work of the authors.

\bibliographystyle{IEEEtran}
\bibliography{mybib}

\end{document}